\documentclass[a4paper,10pt,twoside]{cpc-hepnp}

\usepackage{multicol}
\usepackage{graphicx}
\usepackage{booktabs}
\usepackage{amssymb,bm,mathrsfs,bbm,amscd}
\usepackage[tbtags]{amsmath}
\usepackage{lastpage}
\usepackage{CJK}

\begin{document}
\begin{CJK*}{GBK}{song}

\fancyhead[co]{\footnotesize WANG Yu-Sa et al: Measurements of Charge Transfer Efficiency in a Proton-irradiated Swept Charge Device}

\footnotetext[0]{Received 7 February 2013}

\title{Measurements of Charge Transfer Efficiency in a Proton-irradiated Swept Charge Device\thanks{Supported by the HXMT project}}
\author{%
 WANG Yu-Sa$^{1;1}$\email{wangyusa@mail.ihep.ac.cn}%
\quad YANG Yan-Ji$^{1,2}$ \quad CHEN Yong$^{1}$ \quad LIU Xiao-Yan$^{1,2}$
\\ \quad CUI Wei-Wei$^{1}$ \quad XU Yu-Peng$^{1}$ \quad LI Cheng-Kui$^{1}$ \quad LI Mao-Shun$^{1}$
\\\quad HAN Da-Wei$^{1}$ \quad CHEN Tian-Xiang$^{1}$  \quad HUO Jia$^{1}$ \quad WANG Juan$^{1}$
\\\quad LI Wei$^{1}$ \quad HU Wei$^{1}$ \quad ZHANG Yi$^{1}$ \quad LU Bo$^{1}$ \quad YIN Guo-He$^{1}$
\\\quad  \quad ZHU Yue$^{1}$ \quad ZHANG Zi-Liang$^{1}$ }
\maketitle

\address{%
1~(Key Laboratory for Particle Astrophysics, Institute of High Energy
Physics, Chinese Academy of Sciences (CAS), 19B Yuquan Road, Beijing
100049, China)
\\2~(College of Physics, Jilin University, No.2519, Jiefang Road, Changchun 130023, China)}

\begin{abstract}
Charge Coupled Devices (CCDs) have been successfully used in several low energy X-ray astronomical satellite over the past two decades.
Their high energy resolution and high spatial resolution make them an perfect tool for low energy astronomy, such as formation of
galaxy clusters and environment of black holes. The Low Energy X-ray Telescope (LE) group is developing Swept Charge Device (SCD) for
the Hard X-ray Modulation Telescope (HXMT) satellite. SCD is a special low energy X-ray CCD, which could be read out a thousand times
faster than traditional CCDs, simultaneously keeping excellent energy resolution. A test method for measuring the charge transfer efficiency (CTE) of a prototype SCD has been set up. Studies of the charge transfer inefficiency (CTI) have been performed at a temperature range
of operation, with a proton-irradiated SCD.
\end{abstract}

\begin{keyword}
CCD, SCD, HXMT, LE, CTE, CTI, proton-irradiated
\end{keyword}

\begin{pacs}
29.40.Wk.
\end{pacs}

\begin{multicols}{2}

\section{Introduction}
The Low Energy X-ray Instrument (LE) on the Hard X-ray Modulation Telescope (HXMT) satellite\cite{lab1} is the first
low energy X-ray observation (1.0-15keV) attempt of China. The HXMT mission is scheduled to be launched in early 2015,
with a planned mission duration of 4 years in a 550km circular orbit around the earth. As a part of HXMT,
LE is an X-ray spectrometer
with good full width at half maximum (FWHM) better than 450eV@5.9keV, high time resolution of no more than 1ms and large
detection area (384$cm^2$).

Because the space environment of satellite is filled with protons, neutrons and $\gamma$-rays,
the study of radiation hardness is important for the application of CCD detector in astronomy satellite\cite{lab2}. The LE group
has developing and testing several CCDs for about 6 years. Previous experimental results on proton and neutron irradiation
were reported in\cite{lab3}. The measurements and analysis presented in this paper have been also carried out in the LE group.
This work focuses on an effective experimental measurement method to determine the charge transfer efficiency (CTE)
performed with the proton-irradiated SCD at a test stand, in the Institute of High Energy Physics, Chinese Academy of Sciences.

The particles around the orbit create damage to the SCD materials which leads to defects acting as charge traps in the silicon,
especially the protons in the South Atlantic Anomaly (SAA) can make serious consequences to SCD. The mechanism of creating traps
has been discussed in the literature\cite{lab4}. In a phosphorus-doped devices like SCD, two types of traps are
created\cite{lab5}. A trap can capture one electron in about 1ms and increase dark current,
which makes the electron cloud signal fluctuate apparently.
Finally, this traps result in charge transfer inefficiency and degradation of energy resolution.

LE is in development to undergo series of space environment tests. LE's SCD is working with 100 kHz readout
frequency. In this paper we demonstrate a simple and effective method to determine the CTE with a proton-irradiated SCD.
The method is crucial to check the performance of SCD in environment tests like thermal cycle, proton irradiation and $\gamma$
irradiation.

\section{Test stand and work modes of SCD }
A test stand has been set up with collimator and cooling unit. The temperature range of the cooling unit is from room temperature
down to $-100\,^{\circ}\mathrm{C}$. This temperature has been achieved with cold nitrogen gas by boiling liquid nitrogen. The very
low operating temperature is required to suppress the dark current and keep the SCD sensitive to very low energy signal. The collimator attached to the cooling unit can keep the X-ray react as few pixels as possible. The collimator is machined in a whole
aluminium block by line-cutting. The minimal grid of collimator is about 1.47mm, which is much longer than the size of pixel, 0.1mm. Therefore we stick a 0.5mm copper film on the top of collimator, and punch a hole in the copper film with 0.5mm in diameter.
The hole can keep the X-ray react with even less pixels. On the top of copper film, a 0.5mCi$_{26}^{55}\mathrm{Fe}$ radioactive source
is installed by black 3M tape. The $_{26}^{55}\mathrm{Fe}$ is uniformly distributed in a copper disk by electroplate. As shown in
Fig.~\ref{fig1}, The whole test stand is settled into a vacuum-low temperature tank.
\begin{center}
\includegraphics[width=8cm]{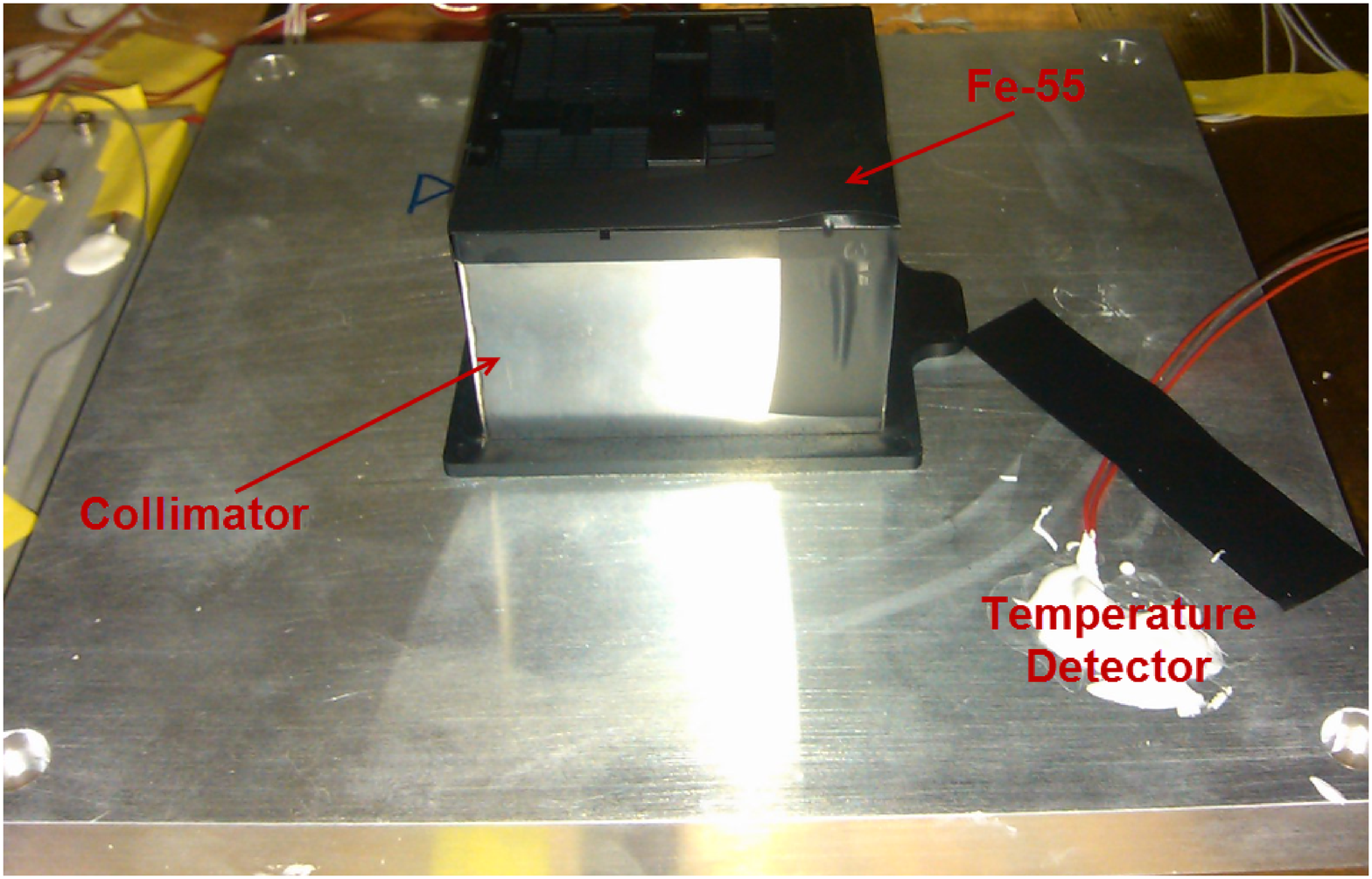}
\figcaption{\label{fig1} Test stand of SCD.}
\end{center}

Under this test stand we can obtain different responses from different pixels through changing work model of SCD in FPGA.
SCD has been created by E2V (English Electric Valve Company Ltd.) and is a relatively large area four-quadrants detector having
400$\mathrm{mm}^{2}$ active area. The active area of four quadrants are covered with 100 ``L" electrodes respectively, as
shown in Fig.~\ref{fig2}. The ``L" electrodes are depicted as the dashed lines without arrow in the figure, the arrow lines stand
for the direction of charge transfer. The charge just needs to be transferred along one dimension because of ``L" electrodes,
so its time for transferring one frame is less than that of the normal X-ray CCDs greatly, about 1ms. The pixel and electrode is
almost same here, but different from two-dimension CCD. One pixel is a ``L" electrodes, the first pixel is only one point, but
the last pixel is a 10+10mm long ``L" electrodes\cite{lab6}.
\begin{center}
\includegraphics[width=6cm]{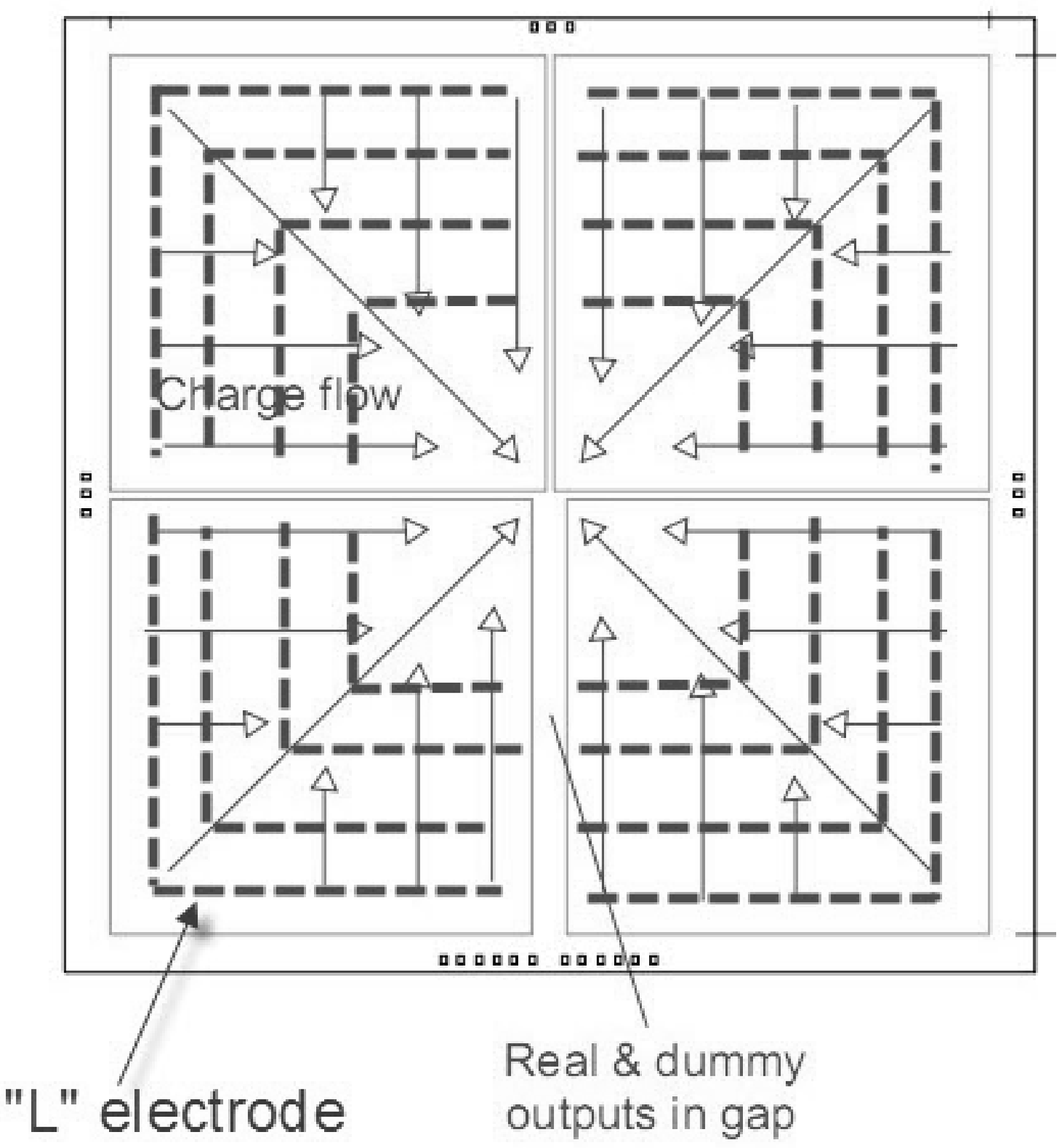}
\figcaption{\label{fig2} Schematic diagram of SCD.}
\end{center}
There are only two work models of SCD because of one dimension readout, the spectrum model and the long exposure mode.
The spectrum mode is a normal readout sequence for shaping spectrum, under which the readout is reset per sequence clock. The spectrum
mode is usually adopted in test for achieving high energy resolution X-ray spectrum and hight time resolution X-ray sources.
In addition to the spectrum mode, the charge packet could be stored at the pixel for many sequence clocks, then whole active area is
read out by the 100 continuous clocks like operation of spectrum mode. This mode is called ``long exposure mode". Through this operation
of charge transfer, we can learn the dark current of every electrode and check whether some electrodes are hot pixel, which generate dark
current over normal level. More importantly, the long exposure mode could show the one dimension position distribution of SCD with
accurate pixel position information.

When the $_{26}^{55}\mathrm{Fe}$ radioactive source is fixed on the copper film with hole, the $_{26}^{55}\mathrm{Fe}$ could irradiate
the SCD through the hole in several pixels (electrodes). Under long exposure mode of SCD, the position distribution of X-ray on the 100
electrodes can be revealed, as shown in Fig.~\ref{fig3}.
\begin{center}
\includegraphics[width=8cm]{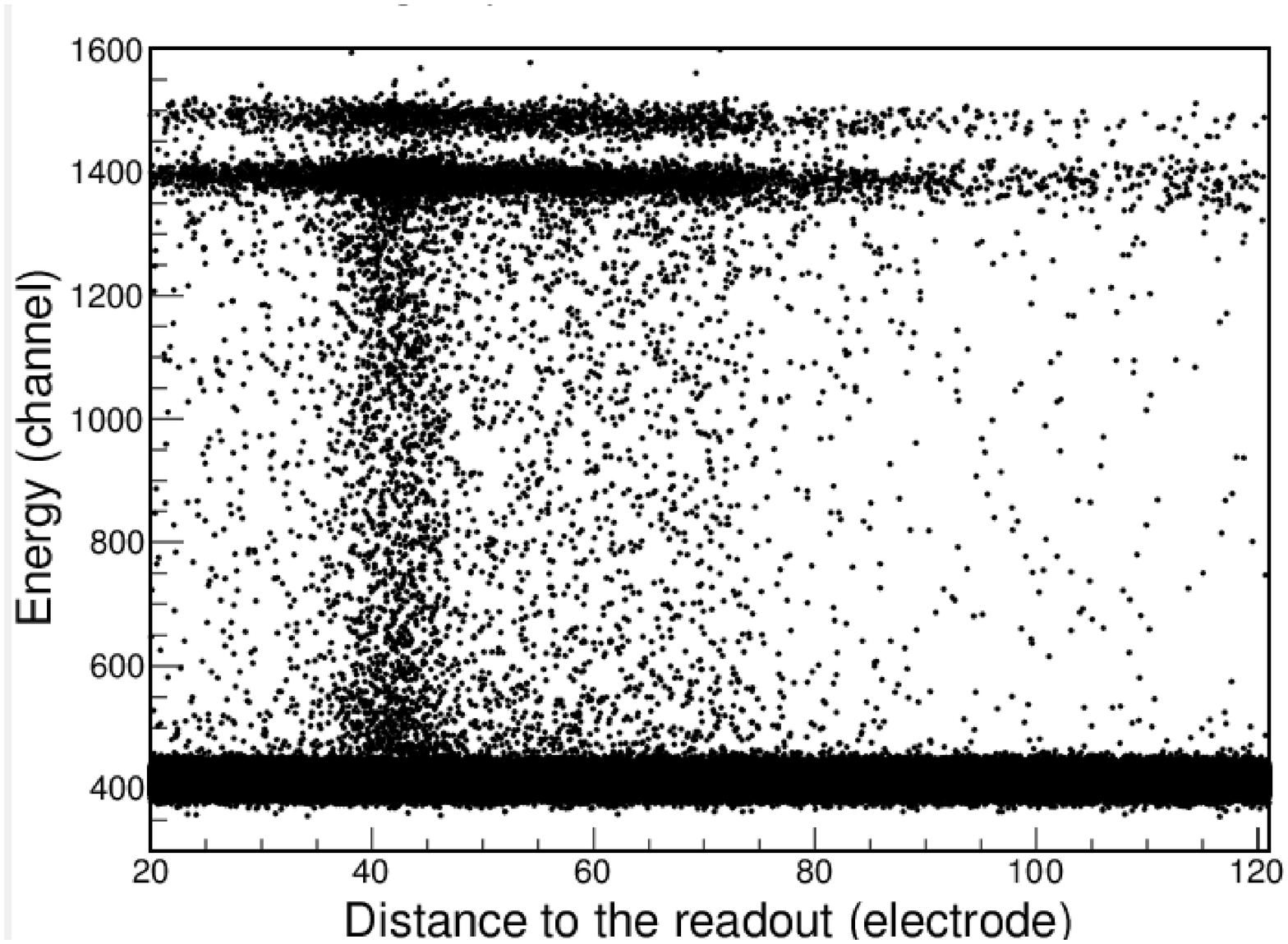}
\figcaption{\label{fig3} The position distribution of X-ray under long exposure mode. The position of the hole is at the about
22th pixel}
\end{center}
The down part of scatter diagram (about 400 channels) is the noise events, the middle and top part of scatter diagram are the
Mn k$\alpha$ (about 1400 channels) and Mn k$\beta$ (about 1500 channels) events. Through this operation the hole position
correspond to pixels is confirmed, knowing the reaction area of X-ray in SCD. Then we change SCD'mode to spectrum mode and acquire
sufficient events to fit the performance parameters at different temperatures. After these two operations of two modes, the response
of one hole is obtained. When we change the $_{26}^{55}\mathrm{Fe}$ radioactive source to the other hole in copper film and do the
same operations like the first hole, the other hole position correspond to pixels and performance parameters at different temperatures
can be confirmed. Since the CTE is used to represent efficiency of the charge transfer from one electrode to another, it could be
determined from the two responses of two holes, reasonably.

\section{Signal measurement of different position}
The signal measurement of two holes is designed to determine the CTE of different pixels. The measurement procedure of the first
hole is as follow. First, we put the $_{26}^{55}\mathrm{Fe}$ radioactive source on the first hole above the 88th pixel from
the readout. Then we acquire the events under long exposure mode and plot the scatter diagram of the first hole. Through
Gauss-Fitting in ROOT the mean position of first hole correspond to SCD is confirmed in Fig.~\ref{fig4}.
\begin{center}
\includegraphics[width=8cm]{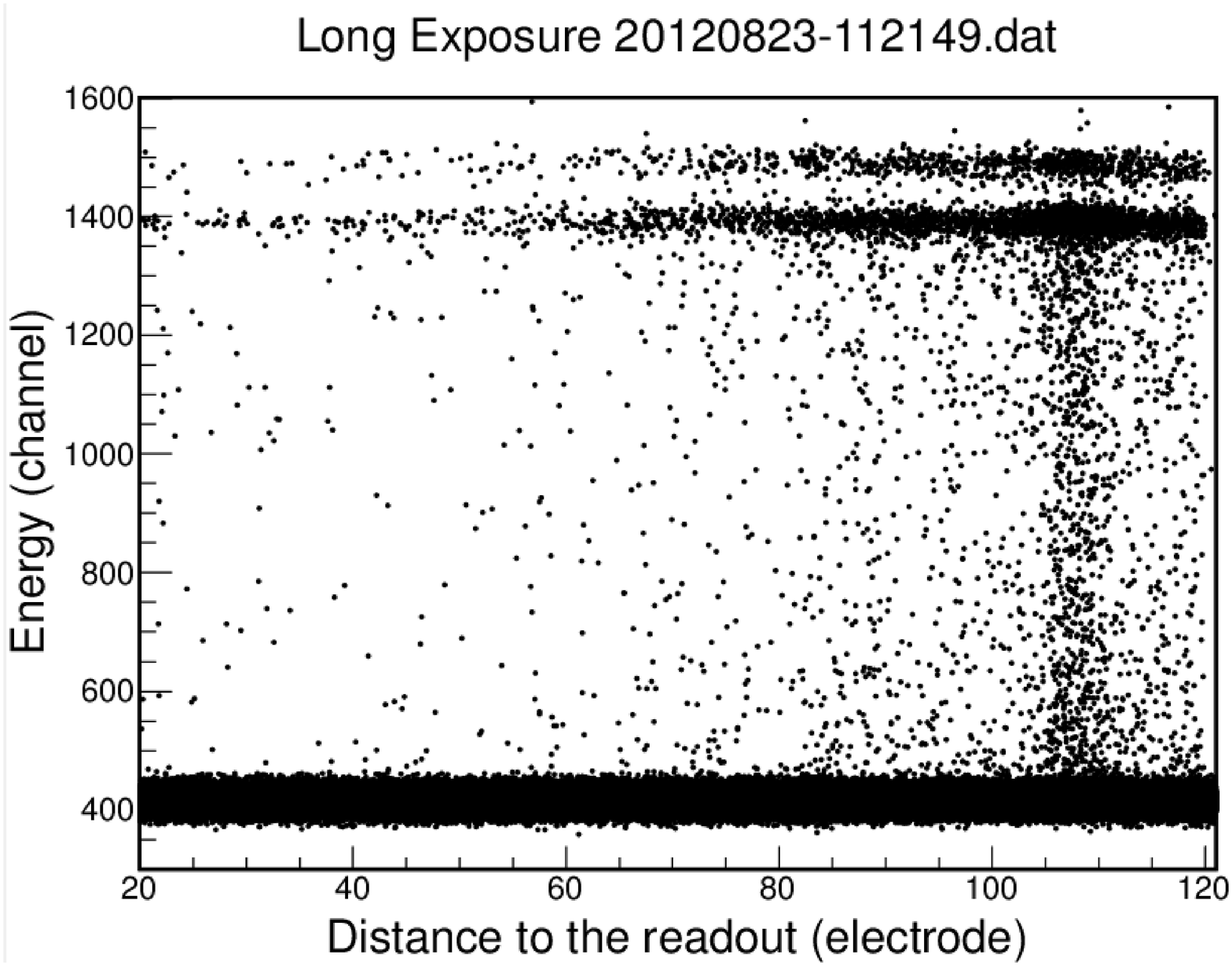}
\figcaption{\label{fig4} The X-ray position distribution of the first hole in a proton-irradiated SCD.
The position of the hole is at the about 88th pixel}
\end{center}
The third step, large mounts of X-ray events are acquired under spectrum mode, with rise in the temperature of SCD. Since the SCD
is working all the time when the temperature is rising slowly, the signals correspond to different temperatures are obtained through
Gauss-Fitting finally. The measurement is completed in vacuum-low temperature tank, low temperature $-100\,^{\circ}\mathrm{C}$,
vacuum $10^{-5}\mathrm{Pa}$. For the second hole the measurement procedure is the same, only needing to move the $_{26}^{55}\mathrm{Fe}$ radioactive source to the second position above the 22th pixel from the readout. Likewise, the second hole correspond to SCD is
confirmed in Fig.~\ref{fig5}, and the signal-temperature curves of two holes are shown in Fig.~\ref{fig6}.
\begin{center}
\includegraphics[width=8cm]{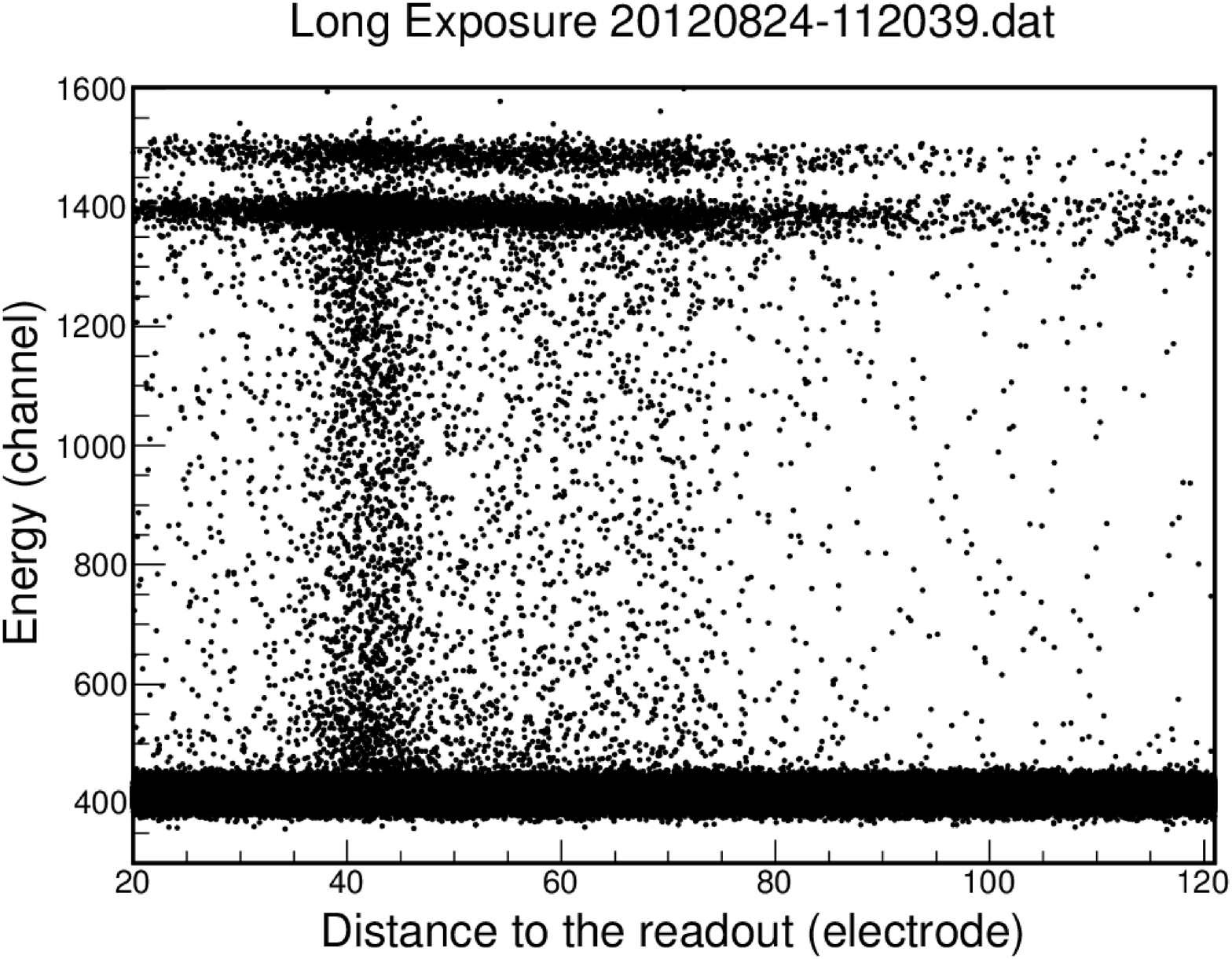}
\figcaption{\label{fig5} The X-ray position distribution of the second hole in a proton-irradiated SCD.
The position of the hole is at the about 22th pixel}
\end{center}
\begin{center}
\includegraphics[width=8cm]{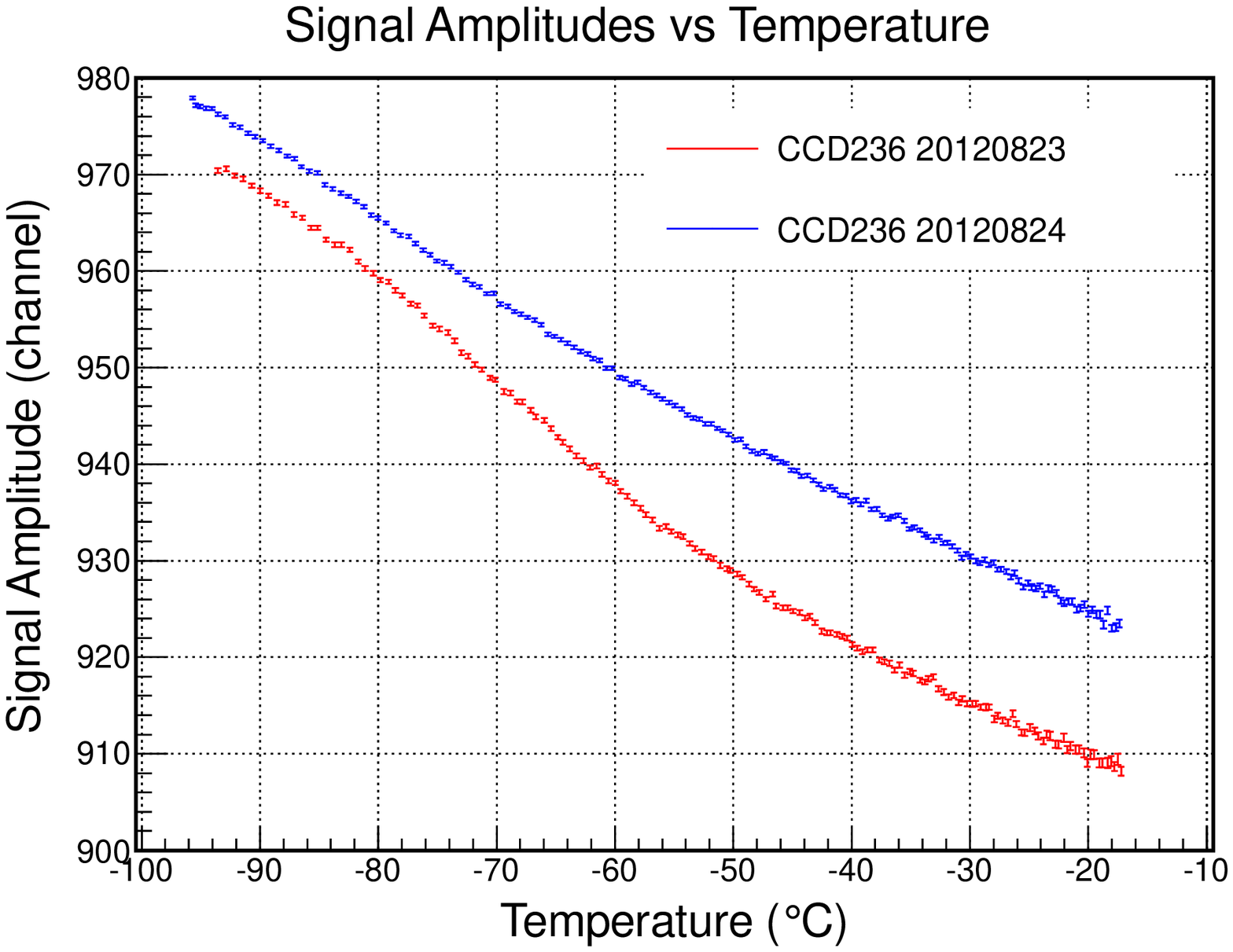}
\figcaption{\label{fig6} The signal-temperature curves of two holes. The bule line is the signal amplitude of the second hole, which is
nearer to the readout than the first hole.}
\end{center}
The signal amplitude of the pixels near to readout is obviously higher than that of far pixels, and the signal-temperature curve of the
far pixels is not linear, with quick decreasing at the temperature range from $-70\,^{\circ}\mathrm{C}$ to $-40\,^{\circ}\mathrm{C}$.
This indicates that the low temperature is helpful to charge transfer of SCD irradiated by proton. Besides, the measurement of signal
amplitude needs to meet the high stability and repeatability requirement. In out test, we keep the test stand almost no change, only
moving the $_{26}^{55}\mathrm{Fe}$ radioactive source slightly. Therefore the two noise peaks in the measurement is nearly the same in
Fig.~\ref{fig7}.
\begin{center}
\includegraphics[width=8cm]{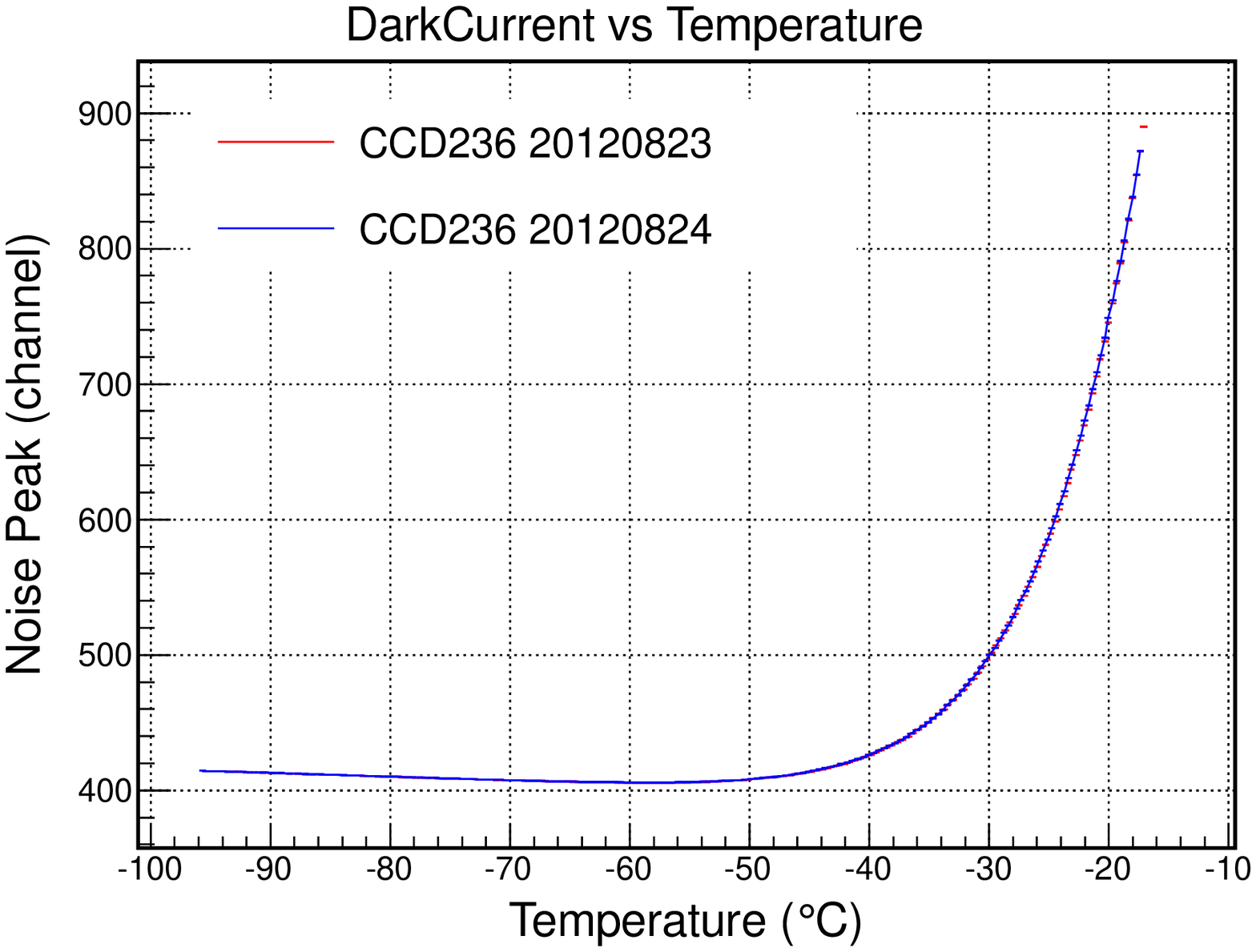}
\figcaption{\label{fig7} The noise peaks of two holes in measurement of CTE.}
\end{center}
The noise peak is the background of signal and eliminated in calculating the finally signal amplitude, which can confirm the
credibility of signal amplitude.

\section{Calculation the CTE}
In the above test, the two signal amplitudes correspond to two holes are determined at the temperature range from about
$-90\,^{\circ}\mathrm{C}$ to $-20\,^{\circ}\mathrm{C}$. When the CTE of every pixel is assumed to be the same, the CTE could be
calculated through formula in Eq.~\ref{eq1} reasonably.
\begin{equation}
\label{eq1}
Amplitude1 = Amplitude2*CTE^{Pixel2-Pixel1}.
\end{equation}
Where the Pixel2 and Pixel1 are the electrode order numbers from the readout for the second and first hole respectively. Similarly, the
Amplitude2 and Amplitude1 stand the two signal amplitudes of two holes. Through the fitting of data, the Pixel2 is about 22, the Pixel1
is about 88. Amplitude1 and Amplitude2 could be obtained from the events acquired under spectrum mode. Through calculation with
Eq.~\ref{eq1}, the CTE is determined in Fig.~\ref{fig8}.
\begin{center}
\includegraphics[width=8cm]{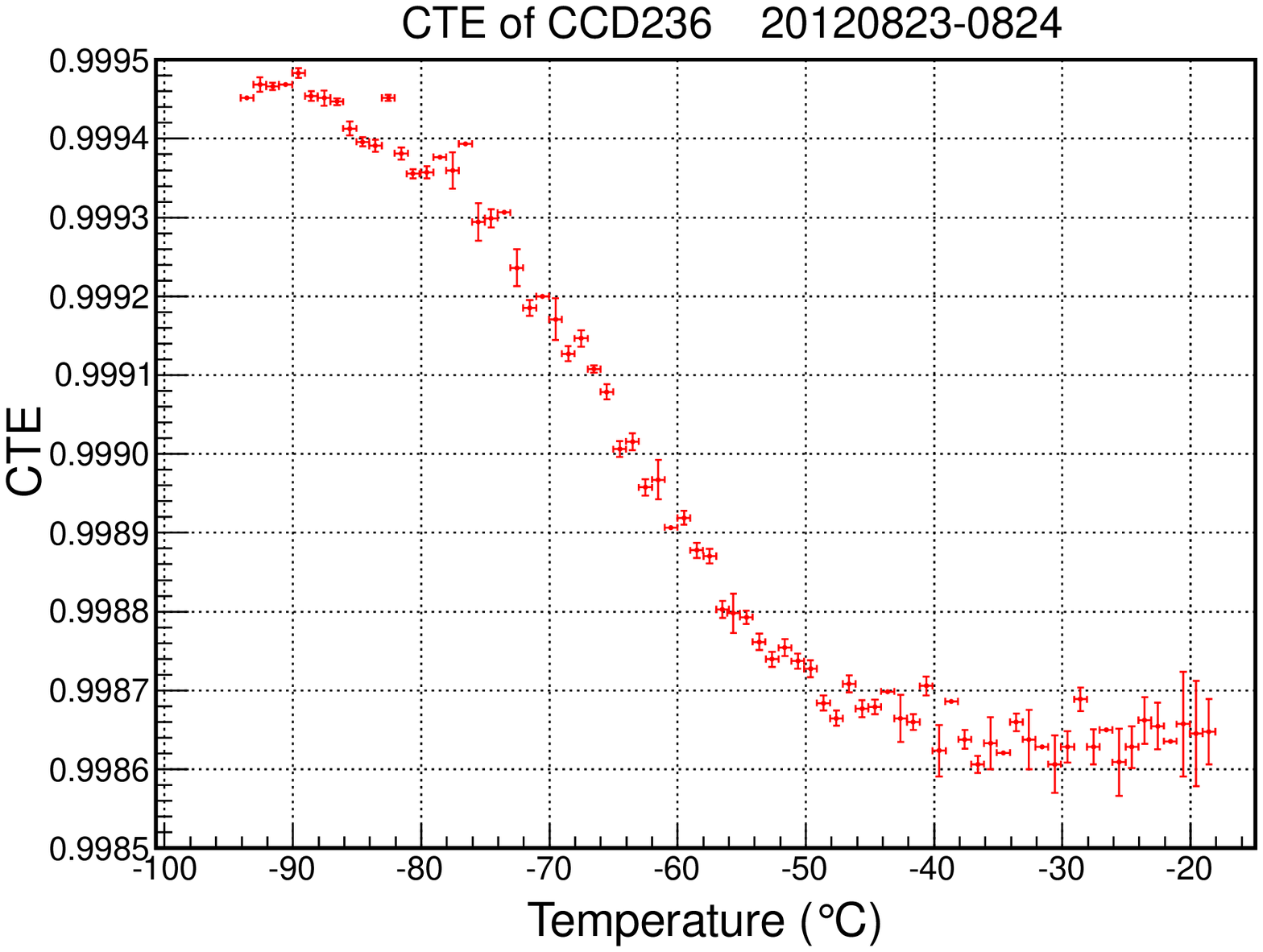}
\figcaption{\label{fig8} The CTE of a proton-irradiated SCD (CCD236).}
\end{center}
As shown in Fig.~\ref{fig8} the CTE is sensitive to temperature, low temperature is helpful to achieve low CTI (Charge Transfer
Inefficiency) for proton-irradiated SCD. This result affirms the importance of thermal control of LE, which could suppress the dark
current on the other hand\cite{lab7}.

\section{Conclusion and outlook}
A proton-irradiated SCD is operated at a range of temperature from $-70\,^{\circ}\mathrm{C}$ to $-40\,^{\circ}\mathrm{C}$ with different
operation modes. The spectrum and energy distribution of a $_{26}^{55}\mathrm{Fe}$ radioactive source are acquired with the SCD. The CTE
is calculated from the comparison of the two signal amplitudes correspond to two holes. The noise peaks are almost same, which shows the
method of CTE measurement reliable. The CTE value is sensitive to high temperature, indicating the necessity of low
temperature. Further CTE measurements with $\gamma$-irradiated SCD are planned. We will also try to measure the CTE of SCD with
more convenient and effective method basing on the integration of dark current under the long exposure mode.

\end{multicols}

\vspace{-2mm} \centerline{\rule{80mm}{0.1pt}} \vspace{2mm}

\begin{multicols}{2}

\end{multicols}

\vspace{5mm}

\clearpage

\end{CJK*}
\end{document}